\documentclass[aps,pra,showpacs,preprint]{revtex4-1}
\usepackage{amsmath}
\usepackage{graphicx}
\usepackage{dcolumn}
\usepackage{bm}
\usepackage{textcomp}
\usepackage{natbib}
\usepackage{upgreek}
\usepackage{enumerate}

\def\bra#1{\mathinner{\langle{#1}|}}
\def\ket#1{\mathinner{|{#1}\rangle}}

\begin{document}
\title{Polarization dependent tuning of the Hanle effect in the ground state of Cs}
	
\author{Harish Ravi}
\author{Mangesh Bhattarai}
\author{Vineet Bharti}
\author{Vasant Natarajan}
\email{vasant@physics.iisc.ernet.in}
	
\affiliation{Department of Physics, Indian Institute of Science, Bangalore-560012, India}

\begin{abstract}
We demonstrate that the Hanle effect can be tuned between magnetically induced absorption (MIA) and magnetically induced transmission (MIT) simply by changing the polarization of the input laser beam. The experiments are done using closed hyperfine transitions of the $ \rm D_2 $ line of ${\rm ^{133}Cs}$ ---$ F_g = 3 \rightarrow F_e = 2 $ and $ F_g =4 \rightarrow F_e = 5 $. The former shows a transformation from MIT to MIA, while the latter shows the opposite behavior. A qualitative explanation based on optical pumping and coherences among the magnetic sublevels of the ground state is borne out by a detailed density-matrix calculation. To increase the coherence time, the experiments are done in a Cs vapor cell with paraffin coating on the walls. The observed linewidth is extremely narrow ($\sim 0.1$ mG) compared to previous work in this area, making this a promising technique for all kinds of precision measurements.
\end{abstract}

\maketitle

\section{Introduction}
Long-lived atomic coherences play an important role in precision measurements. For example, in the phenomenon of nonlinear magneto-optic rotation (NMOR), a laser field is used to align the atom by inducing coherences among the magnetic sublevels of the ground state. NMOR has important applications in sensitive magnetometry \cite{BGK02}, and the search for a permanent electric dipole moment (EDM) in an atom \cite{RCN11}. Another example is an atomic clock, where two laser beams which differ in frequency by the clock transition are used to put the atom in a dark non-absorbing state. The phenomenon of coherent population trapping (CPT) \cite{ARI96} is then used to get a narrow resonance at exactly the clock frequency. Both these kinds of experiments gain from having a vapor cell filled with buffer gas \cite{BNW97}, or anti-relaxation (paraffin) coating on the walls \cite{BOB66,RMN15}. This reduces the spin relaxation rate and increases the coherence time, thus resulting in a narrower linewidth.

In this work, we demonstrate the tuning of the Hanle effect from magnetically induced transparency (MIT) to magnetically induced  absorption (MIA), and \textit{vice versa}, in the ground state of Cs atoms \footnote{The use of the term MIT (which is standard terminology in the field) is not to imply that there is more light after the medium than before. Rather, MIT means that there is reduced absorption compared to an unpolarized medium.}. The atoms are contained in a room temperature vapor cell with paraffin coating on the walls. The change is achieved by changing the polarization of the light from linear to circular. Our work is different from previous work in this field because of the following reasons.
\begin{enumerate}

	\item MIT and MIA have been studied before, but only separately \cite{BTY05,DGM08,KIM12,RAP10}
	
	\item Only the transformation from MIA to MIT (using transitions in the $ \rm {D_1} $ line of Rb) has been studied before \cite{YLB10,MOK14}.
	
	\item The effect of a single kind of polarization in the $ \rm {D_1} $ line of Cs has been studied, both linear \cite{BRW12} and circular \cite{CAW11}.
	
\end{enumerate}
In the present work, we demonstrate for the first time both kinds of transformation---MIT to MIA and MIA to MIT---in the same species of atoms. The experiments are done using closed hyperfine transitions in the $ \rm {D_2} $ line of $ \rm {^{133}Cs} $. Using closed transitions has the advantage of making the theoretical analysis easier, because atoms can not go out of the levels being considered.


The transitions used are $ F_g =3 \rightarrow F_e = 2 $ and $ F_g = 4 \rightarrow F_e = 5 $. The former shows a transformation from MIT to MIA as the polarization is changed from linear to circular, while the latter shows the opposite behavior. The explanation for this is that population redistribution due to optical pumping is different in the two cases, because the number of sublevels is smaller in the excited state compared to the one in the ground state for the first case while it is greater in the second case. Optical-pumping effects are best understood in closed transitions like the ones used here. This qualitative explanation is borne out by a detailed  density-matrix calculation, which produces spectra that are in good agreement with the experimental results. As expected, paraffin coating increases the lifetime of spin coherences, which results in extremely narrow linewidth for the spectra---only about 0.1 mG in all cases. Furthermore, in contrast to previous work, a small transverse magnetic field (of order 0.1 mG) is needed for the circular polarization effect to work.

\section{Experimental details}

\begin{figure}
	\centering
	\includegraphics[width=0.8\textwidth]{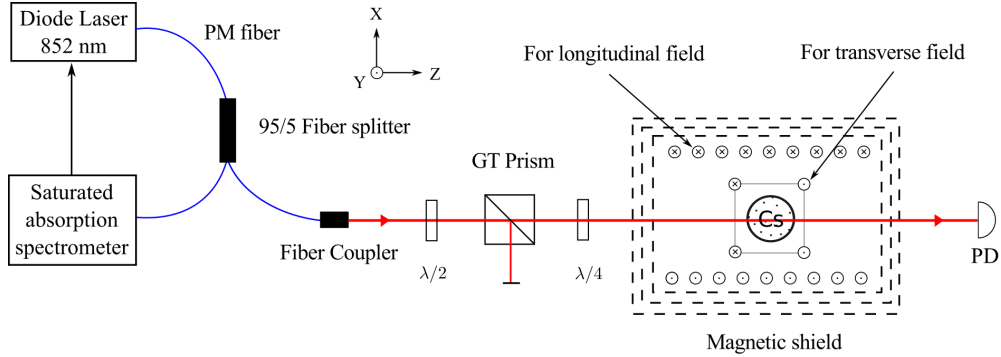}
	\caption{Schematic of the experiment. Figure key: PM fiber -- polarization-maintaining fiber; $\lambda /2$ -- half-wave retardation plate; GT prism -- Glan-Taylor prism; $\lambda /4$ -- quarter-wave retardation plate; PD -- photodiode.}
	\label{fig:Hanleexpt}
\end{figure}

The experimental set up is shown schematically in Fig.\ \ref{fig:Hanleexpt}. The laser beam is derived from a commercial diode laser system (Toptica DL Pro) operating near the 852 nm $\rm D_2 $ line of ${\rm^{133}Cs}$. The laser output comes out of a single-mode polarization-maintaining fiber. The fiber goes into a 95/5 power splitter, with $ 5\% $ of the power fed to a Toptica compact Doppler-free spectroscopy (CoSy) unit. This provides a saturated-absorption signal so that the laser can be locked to any hyperfine transition. The remaining $ 95\% $ of the power is coupled to free space using a fiber coupler. The output goes into a Glan-Taylor prism, which serves the dual purpose of controlling the power (using a half-wave retardation plate), and making the polarization almost perfectly linear (because the extinction for the orthogonal polarization is better than $10^5 $). A quarter-wave plate after the prism is used to control the ellipticity of the beam.

The laser beam after the free-space coupler has a Gaussian intensity distribution with $ 1/e^2 $ diameter of 3 mm and power adjusted to $ 100 $ \textmu W. This gives a maximum intensity at the center of the beam as 2.83 mW/cm$^2$, which corresponds to a Rabi frequency of $ 1.2 \, \Gamma $ from the saturation intensity $I_{\rm sat}$ of 1.01 mW/cm$^2$. The beam goes into a spherical Cs vapor cell with 75 mm diameter and paraffin coating on the walls. The cell is inside a 3-layer magnetic shield with a shielding factor of better than $ 10^4 $. The required longitudinal field for the Hanle effect is obtained by placing the cell inside a solenoid coil. The solenoid is wound on a plastic form of 190 mm diameter. It consists of 1800 turns of 0.35 mm magnet wire wound tightly over a length of 640 mm. Transverse fields are applied using a set of coils on a square form of 106 mm diameter. The power of the laser beam after it passes through the cell is measured using an amplified photodiode.

\section{Experimental results}

${\rm^{133}Cs}$ has two values of the total angular momentum $ F $ in the ground state---3 and 4. Therefore there are two closed transitions in the $\rm D_2 $ line: $ 3 \rightarrow 2 $ and $ 4 \rightarrow 5 $.

\begin{figure}
	\centering
	\includegraphics[width=0.5\columnwidth]{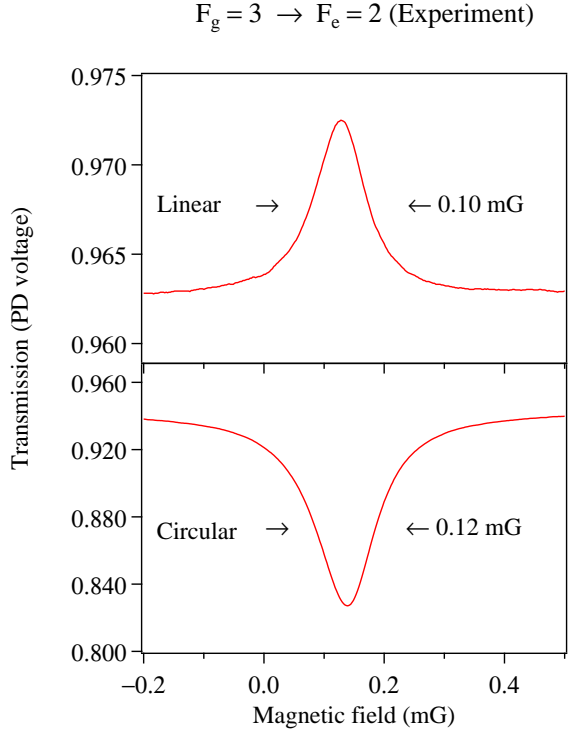}
	\caption{Experimental results for the $3\rightarrow 2 $ transition. Transmission as a function of applied longitudinal B field shows transformation from MIT for linear polarization to MIA for circular polarization.}
	\label{fig:3to2}
\end{figure}

Experimental results for the Hanle effect on the $ 3 \rightarrow 2 $ transitions are shown in Fig.\ \ref{fig:3to2}. The resonance as the longitudinal B field is scanned is extremely narrow  because of the use of a paraffin-coated vapor cell. The resonance peak shows increased transmission (or MIT) for linear polarization, which transforms to reduced transmission (or MIA) for circular polarization. This implies that the medium can be tuned between \textit{slow} and \textit{fast} light simply by changing the polarization of the light, because the sign of the dispersion depends on whether absorption is enhanced or reduced \cite{BHN15}.

The measured photodiode voltage is scaled to reflect the actual absorption through the cell. The scaled photodiode signal accurately reflects probe absorption, both off resonance (high magnetic fields), and the percentage change near resonance. It is seen that it is 10 times higher for the circular case compared to the linear one, which necessitates averaging over 100 traces for the linear case.

The resonances are not centered at zero field because of the presence of a small residual B field inside the shield---both longitudinal and transverse. The longitudinal component causes a shift from zero. The transverse component is required for the circular polarization effect to work (as explained in the \textbf{theoretical analysis} section); this is different from other Hanle experiments where the transverse field is either unimportant or actively nulled. The linewidth obtained in both cases is about 0.1 mG, and is limited by spin relaxation processes. This is the linewidth obtained in a paraffin coated cell which increases by a factor of 100 in a normal cell.

\begin{figure}
	\centering
	\includegraphics[width=0.5\columnwidth]{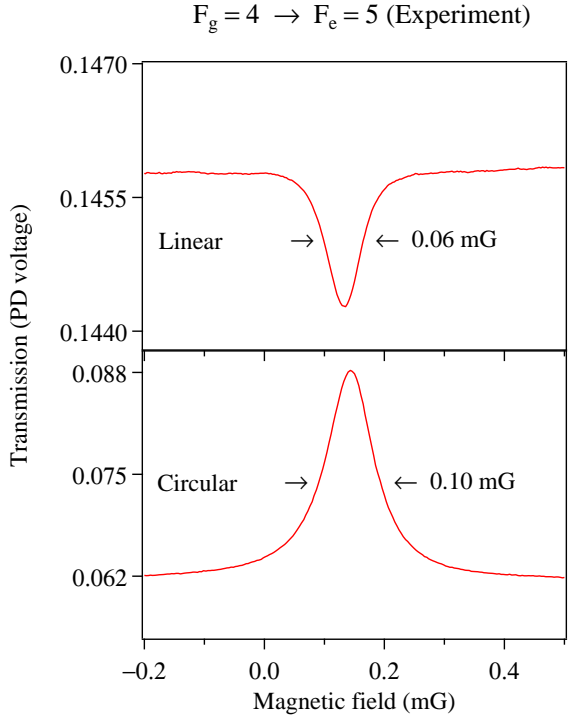}
	\caption{Experimental results for the $ 4\rightarrow 5 $ transition. Transmission as a function of applied longitudinal B field shows transformation from MIA for linear polarization to MIT for circular polarization.}
	\label{fig:4to5}
\end{figure}

Experimental results for the corresponding closed transition starting from the upper ground level, namely $ 4 \rightarrow 5 $ are shown in Fig.\ \ref{fig:4to5}. As before, the percentage absorption for the linear case is much smaller and the curve is obtained after averaging over 100 traces. In addition, the resonances are not centered at zero field because of the presence of a residual longitudinal field.

\section{Theoretical analysis}
In order to verify that the experimental results match theoretical understanding, we have done a theoretical calculation of the same. The calculations were carried out using the atomic density matrix (ADM) package for Mathematica written by Simon Rochester (can be downloaded from http://rochesterscientific.com/ADM/). It solves numerically the following time evolution equation for the density-matrix elements involved:
\begin{equation}
\frac{d\rho}{dt} = \frac{1}{i\hbar}[\rho,H] + \rm{relaxation \, and \, repopulation \, terms}
\end{equation}
The calculation takes as input---(i) the two $ F $ values for the transition, (ii) the light intensity, and (iii) the polarization ellipticity. Experimental parameters of incident power of 100 \textmu W in a size of 3 mm, a small transverse field of 0.1 mG, and an interaction length corresponding to the cell size of 75 mm were taken. The atomic number density was taken to be $ 3 \times 10^{10} $ atoms/cc, which corresponds to a vapor pressure of 1 \textmu torr (corresponding to Cs vapor at room temperature). The intensity was assumed to decay exponentially due to absorption along the beam path. 

The simulation assumed two regions---labeled A and B---with different Hamiltonians, because both the light and magnetic field are present in region A while only the magnetic field is present in region B. The atoms relax from region A to region B (transit-time relaxation) at a rate of 50 kHz, which corresponds to the most-probable velocity going across the beam size of 3 mm. The atoms come back from region B to region A at a rate of 2 kHz, determined roughly by the ratio of cell volume to interaction volume. Finally, the relaxation rate at which coherences are lost (due to collisions with the cell wall) is taken to be 25 Hz, chosen to match the linewidth seen in a paraffin-coated vapor cell in our related NMOR experiments \cite{RCN11}. The spectrum is Doppler-averaged over the Maxwell-Boltzmann velocity distribution appropriate for Cs atoms at room temperature.

The algebraic expressions given in Ref.\ \cite{BRW12} for the circular case and Ref.\ \cite{CAW11} for the linear case can be used to understand our experimental results, but only qualitatively. This is because the expressions use many approximations like low intensity ($I \ll I_{\rm sat}$) and a single velocity (no Doppler averaging). In addition, the constants used are different from the experimental ones we have. As a consequence, the Lorentzian lineshape is reproduced for the circular case; but it predicts a splitting for the linear case, which is not observed experimentally. These results are shown in Appendix A.

By contrast, the numerical package that we use allows for a more realistic simulation with transit-time relaxation and wall-collision relaxation rates (which do not change with laser power, atomic number density, etc.), near saturation intensity, and Doppler averaging. In order to improve the understanding of the ADM package, we present an analysis of simplified transitions in the Appendix B. All the details are the same except that the sublevel structure is simplified to deal with $ 1 \rightarrow 0 $ and $ 1 \rightarrow 2 $ transitions.

Magnetic sublevel structure for the two final transitions is shown in the figures below. In all cases, the quantization axis is along the polarization direction, i.e.\ perpendicular to the direction of propagation for linear polarization, and along the direction of propagation for circular polarization. The selection rule for sublevel transitions coupled by linear polarization is $ \Delta m = 0 $. Similarly, the selection rule for right circular polarization ($ \sigma^+ $) is $ \Delta m = +1 $, while the one for left circular polarization ($ \sigma^- $) is $ \Delta m = -1 $.

\subsection{$ F_g = 3 \rightarrow F_e = 2 $ \rm {transition}}
\begin{figure}
	\centering
	\includegraphics[width=0.5\columnwidth]{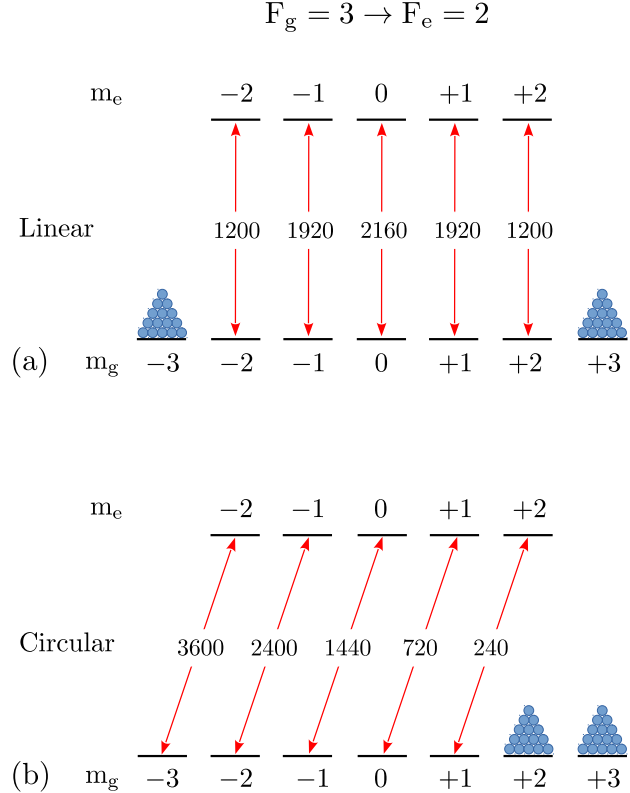}
	\caption{Magnetic sublevels of the $ 3 \rightarrow 2 $ transition in the $\rm D_2 $ line of $^{133}$Cs. The quantization axis is along the direction of polarization. Population distribution after optical pumping and relative strengths for transitions coupled by two kinds of polarizations are shown. (a) Linear. (b) Circular.}
	\label{fig:3to2populations}
\end{figure}

Using the above rules, we can see that the steady state population distribution (after optical pumping) for the $ 3 \rightarrow 2 $ transition will evolve from equal populations in all the sublevels to what is shown in Fig.\ \ref{fig:3to2populations}. With linear polarization, shown in part (a) of the figure, all the population ends up in the $ m_{F_g} = -3 $ and $ +3 $ sublevels. This is because there are no upper-state sublevels to which it can be excited. From the relative strengths of transitions coupled by this polarization (determined by the respective Clebsch-Gordan coefficients), all the population gets optically pumped to non-absorbing sublevels.

Similarly, the distribution after optical pumping with right circular polarization, shown in part (b) of the figure, will end up in the $ m_{F_g} = +2 $ and the $+3$ sublevels. This is because as before there are no upper-state sublevels to which they can be excited. From the relative strengths of transitions, the population is again driven to non-absorbing sublevels. The same argument shows that $ \sigma^- $ light will optically pump atoms to non-absorbing $ m_{F_g} = -2 $ and $ -3 $ sublevels.

One way to think about the effect of optical pumping is to consider that the light induces a magnetic moment in the direction of the quantization axis, which then precesses around a transverse magnetic field. This will result in redistribution of the population into absorbing states in the presence of such a transverse field. By definition, a longitudinal B field (as required in the Hanle effect) is transverse to linear polarization. However, the degree of population redistribution, and hence overall absorption, will increase as the value of the field is increased. This shows that the absorption will be minimum at $ B_{\ell} = 0 $, and the resonance peak in the Hanle effect will show MIT.

Now we consider what happens with circular polarization. In this case, an additional transverse field is required because the induced moment is in the longitudinal direction. The transverse field will cause population redistribution into absorbing levels, but its effect will be progressively smaller as the magnitude of the longitudinal field is increased. This shows that the absorption will be maximum at $ B_{\ell} =0 $, and the resonance peak in the Hanle effect will show MIA.

\begin{figure}
	\centering
	\includegraphics[width=0.5\columnwidth]{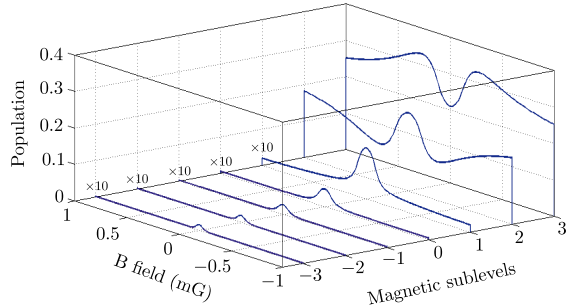}
	\caption{Populations in the different magnetic sublevels of the $F_g = 3 $ ground state as a function of B field strength.}
	\label{fig:F3circular}
\end{figure}

In order to see population redistribution in the presence of a B field transverse to the quantization axis, we show in Fig.~\ref{fig:F3circular} the populations in the various magnetic sublevels of the $F_g = 3 $ state as a function of B field. The curves are shown for circular polarization. The case of linear polarization (not shown) is similar, but complicated by the choice of quantization axis.

\begin{figure}
	\centering
	\includegraphics[width=0.5\columnwidth]{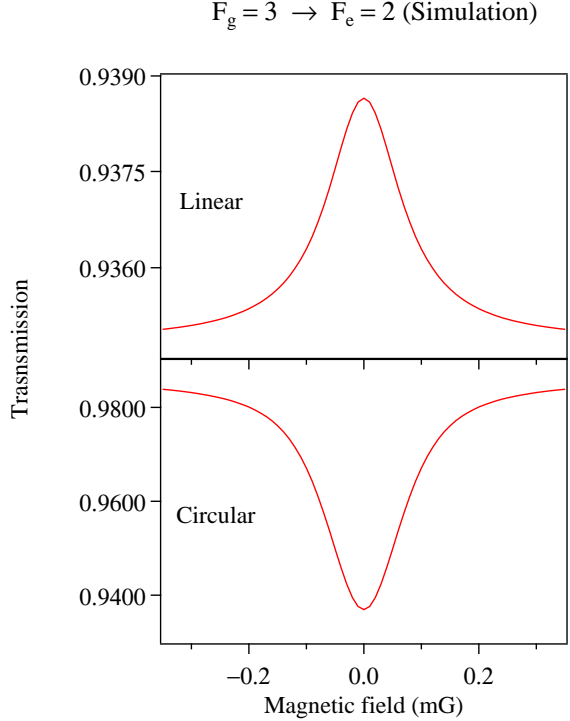}
	\caption{Results of simulation for the $ 3 \rightarrow 2 $ transition showing transformation from MIT to MIA, exactly as seen in the experimental data.}
	\label{fig:3to2sim}
\end{figure}

The results of the simulation, shown in Fig.\ \ref{fig:3to2sim}, are consistent with the qualitative discussion above. The most interesting thing to note is that it reproduces the observed transformation from MIT for linear polarization to MIA for circular polarization. The calculated linewidth is close to the experimental one. The simulation does not reproduce the percentage change in each case, mainly because the simplistic calculation not take into account velocity-changing collisions, and assumes that the intensity is uniform across the beam.

\subsection{$ F_g = 4 \rightarrow F_e = 5 $ \rm {transition}}
\begin{figure}
	\centering
	\includegraphics[width=0.5\columnwidth]{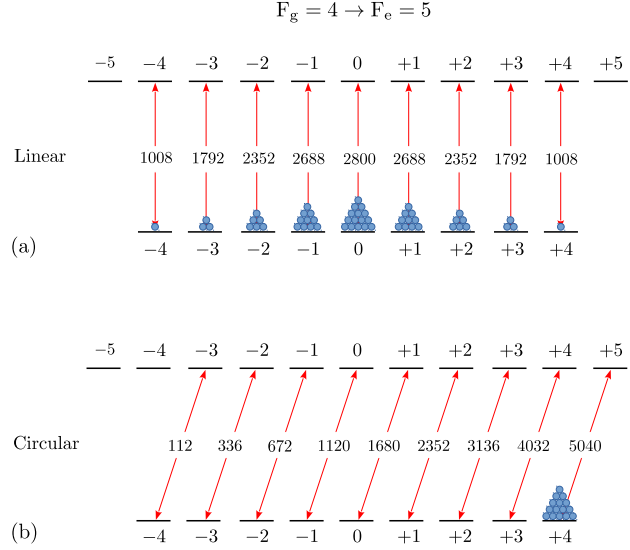}
	\caption{Magnetic sublevels of the $ 4 \rightarrow 5 $ transition in the $\rm D_2 $ line of $^{133}$Cs. The quantization axis is along the direction of polarization. Population distribution after optical pumping and relative strengths for transitions coupled by two kinds of polarizations are shown. (a) Linear. (b) Circular.}
	\label{fig:4to5populations}
\end{figure}

We now consider magnetic sublevels of the $ 4 \rightarrow 5$ transition. The population distribution after optical pumping is shown in Fig.\ \ref{fig:4to5populations}. With linear polarization, shown in part (a) of the figure, maximum population is in the $ m_{F_g} = 0 $ sublevel, and reaches a minimum at the $ m_{F_g} = \pm 4 $ extreme sublevels. This is because the probability of decay from the corresponding upper-state sublevels has this kind of distribution. From the relative transition strengths shown in the figure, this population distribution is going to give maximum absorption. Similarly, optical pumping by right circular polarization ($ \sigma^+ $ light) will result in all the population ending up in the $ m_{F_g} = +4 $ sublevel. The relative transition strengths show that this sublevel is maximally absorbing, therefore as for the linear case absorption will be a maximum when the field is zero.

\begin{figure}[!h]
	\centering
	\includegraphics[width=0.5\columnwidth]{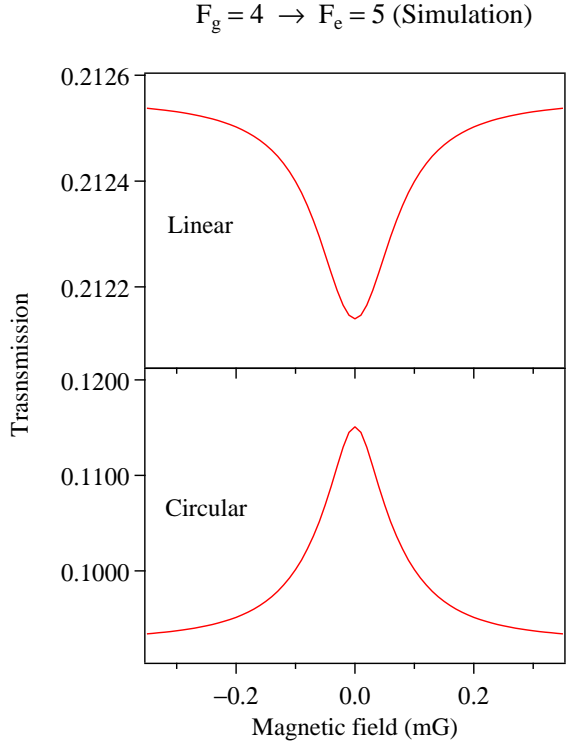}
	\caption{Results of simulation for the $ 4 \rightarrow 5 $ transition showing transformation from MIA to MIT, exactly as seen in the experimental data.}
	\label{fig:4to5sim}
\end{figure}

In the presence of a field transverse to the quantization axis, the population distribution will change and hence the absorption will decrease. Our qualitative analysis proceeds as before. 
\begin{enumerate}
	\item For linear polarization, the longitudinal field used in the Hanle effect will result in progressively smaller absorption. Hence, absorption will be a maximum at $ B_{\ell} =0 $ and the resonance will show MIA.

\item For circular polarization, a transverse field is required because the quantization axis is in the longitudinal direction. As discussed for the other transition, the effect of the transverse field will become progressively smaller as the magnitude of the longitudinal field is increased. Therefore, absorption will be minimum at $ B_{\ell} = 0 $, and the resonance peak will show MIT.
\end{enumerate}

The results of the simulation are shown in Fig.\ \ref{fig:4to5sim}. The results are consistent with the qualitative discussion given above. The detailed calculation reproduces MIT and MIA in each case, and the experimentally observed linewidth. Importantly, it reproduces the observation that the transformation from MIA to MIT is opposite to what was seen for the other transition. As before, the simulation does not reproduce the percentage change as seen in the experimental data, for the same reasons as discussed before.

\section{Conclusions}
In summary, we have demonstrated that the Hanle effect can be tuned between magnetically induced absorption and magnetically induced transmission by changing the ellipticity of polarization of the input laser beam. The experiments are done using the two closed hyperfine transitions of the $\rm D_2 $ line of ${^{133}\rm Cs} $: $F_g = 3 \rightarrow F_e = 2 $ and $ F_g = 4 \rightarrow F_e = 5 $. The former shows a transformation from MIT to MIA when the polarization is changed from linear to circular, while the latter shows transformation from MIA to MIT. The experiments gain by having paraffin coating on the walls of the vapor cell since it increases the relaxation time of coherences among magnetic sublevels. The linewidth obtained in such a cell is of order 0.1 mG, making the technique a good candidate for sensitive magnetometry.

The experimental observations can be understood qualitatively based on optical pumping among the magnetic sublevels of the ground state. A detailed density-matrix calculation using the levels involved bears out the above qualitative explanation. This ability to tune the sign of the dispersion opens up new applications in areas such as slow light and quantum-information processing.

\acknowledgments
This work was supported by the Department of Science and Technology, India. The authors thank Karthik Selvan and Abhilash Y D for help with the experiments; and S.\ Raghuveer for help with the manuscript preparation.
\section*{Appendix A}
In this appendix, we will see how well the algebraic expressions given in Refs. \cite{BRW12} and \cite{CAW11} match our experimental data.

\begin{figure}[h!]
	\centering
	(a) \includegraphics[width=.5 \textwidth]{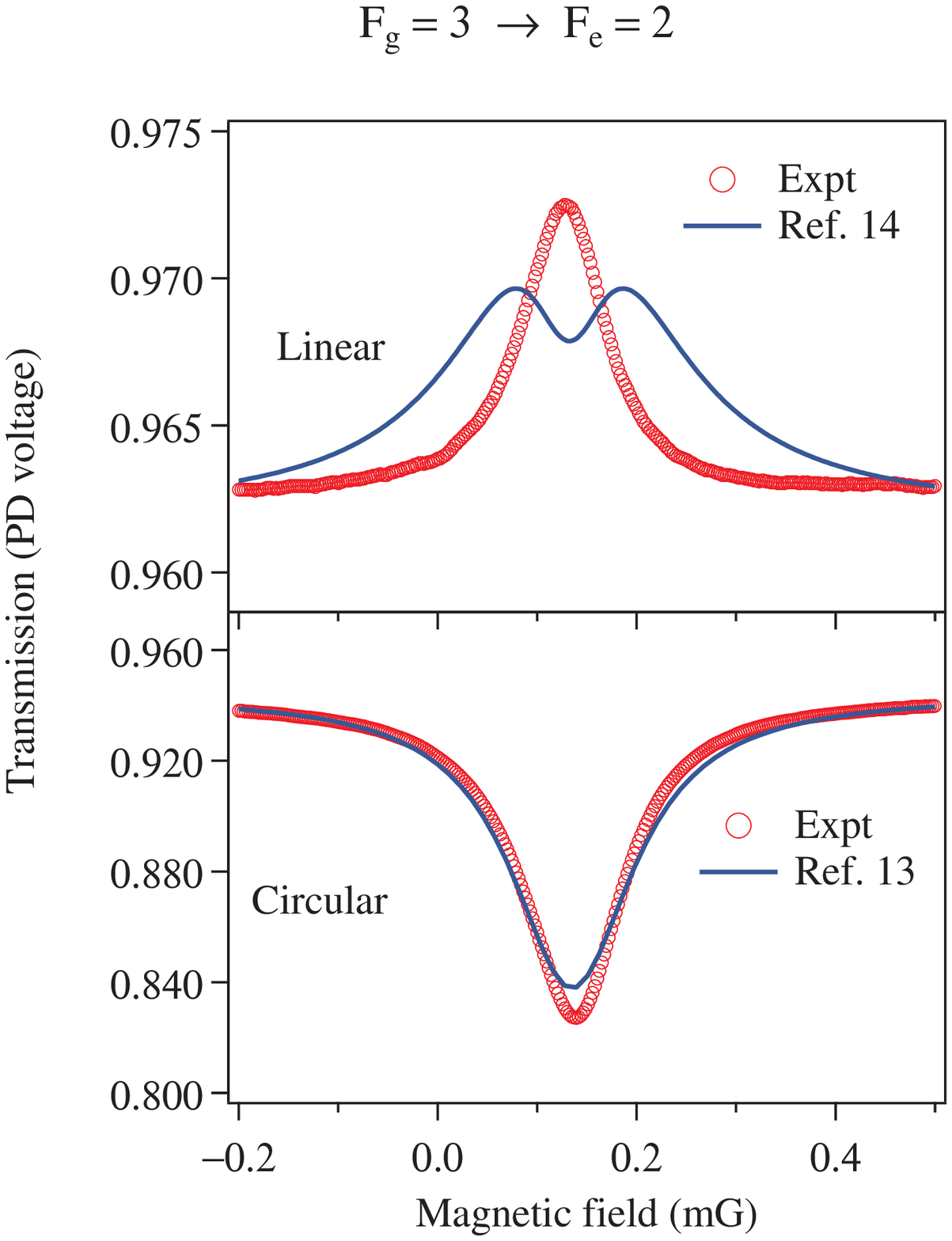}\\
	(b) \includegraphics[width=.5 \textwidth]{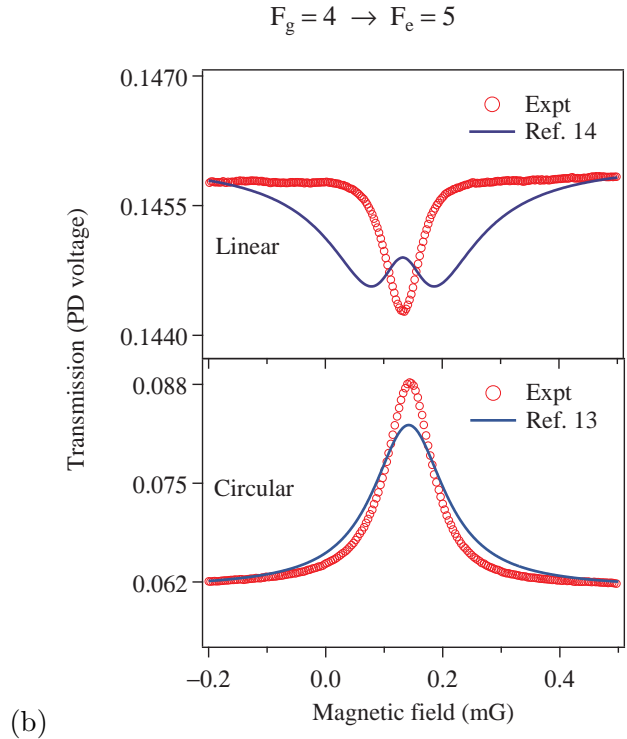}
	\caption{Comparison of experimental data with the results from algebraic expressions given in Refs. \cite{BRW12}, and \cite{CAW11}.  (a) Lower transition. (b) Upper transition.}
	\label{fit}
\end{figure}

For the circular case \cite{BRW12}, the prediction is a single Lorentzian peak with width determined by the transverse B field. The calculation is done with a field of 0.075 mG, which is close to the experimental value and the value used in the ADM package. For the linear case \cite{CAW11}, a transverse B field of 0.075 mG results in a splitting into two peaks, probably because the paraffin coating is neglected. A comparison of the predicted lineshape and experimental data for the different cases is shown in the Fig.\ \ref{fit}.

\section*{Appendix B}
We have computed the Hanle resonances by solving density matrix equations for $F_g = 1 \rightarrow F_e = 0$ and $F_g = 1\rightarrow F_e = 2$ transitions. These transitions are the simplest case for $F_g  \rightarrow F_e = F_g-1$ and $F_g  \rightarrow F_e = F_g+1$ transitions. The time evolution of the system is given by using density matrix analysis for two different regions labeled A and B---with different Hamiltonians, because both the light and magnetic field are present in region A while only the magnetic field is present in region B. The time dependence of density matrix for two regions is given by following set of coupled equations \cite{ROC10}:
	\begin{eqnarray}
	\begin{aligned}
	\dot{\rho}^A&=\dfrac{-i}{\hbar}\left[(H_{int}+H_B), \rho^A\right]-\dfrac{1}{2}\left\{\Gamma,\rho^A \right\}\\
	&-\gamma_{t1}\rho^A+\Lambda^A+\gamma_{t2}\rho^B \\
	\dot{\rho}^B&=\dfrac{-i}{\hbar}\left[H_B, \rho^B\right]-\dfrac{1}{2}\left\{\Gamma,\rho^B \right\}\\
	&+\gamma_{t1}\rho^A+\Lambda^B-(\gamma_{t2}+\gamma_w)\rho^B + \dfrac{\delta^{g.s.}}{2F_g+1}\gamma_w
	\label{mastereqns}
	\end{aligned}
	\end{eqnarray}
	
	\begin{figure}[ht]
		\centering
		\includegraphics[width=.6 \textwidth]{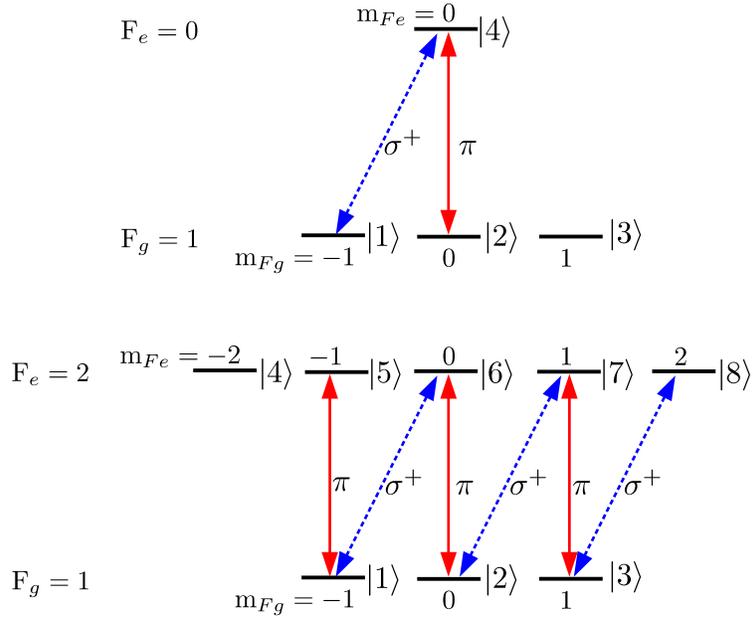}
		\caption{Coupling between magnetic sub-levels and light fields for $F_g =1 \rightarrow F_e =0$ and $F_g =1 \rightarrow F_e =2$ transitions.}
		\label{levelsforsim}
	\end{figure}
	
Here, $H_{int}$ is the interaction Hamiltonian for light-atom interaction and $H_B$ describes the coupling with constant B-fields, $\Gamma$ is the relaxation matrix for spontaneous emission, $\Lambda^{A,B}$ are repopulations in regions A and B, $\gamma_{t1}$ is the transit relaxation from region A to B and $\gamma_{t2}$ is the transit relaxation from region B to A, and $\delta^{g.s.}$ is the identity matrix for ground state. In the case of linear polarization, the quantization axis is chosen along the polarization of light, the longitudinal field along propagation of light and  residual field perpendicular to both. For $\sigma^+$-polarization, the quantization axis and longitudinal field are along the direction of propagation of light and residual field is perpendicular to it.

The magnetic sub-levels for $F_g =1 \rightarrow F_e =0$ and $F_g =1 \rightarrow F_e =2$ transitions are shown in Fig.\ \ref{levelsforsim}. In this figure, the $\pi-$polarized transitions are shown with solid lines and $\sigma^+ -$polarized transitions are shown with dashed lines. The $H_{int}$ for these transitions is described below:
\begin{center}
		\begin{tabular}{|c|c|}
			\hline  \parbox{4em}{\vspace{ .1cm}\vspace{ .1cm}}& $\bf{H_{int}}$  \\ 
			\hline \parbox{4em}{\centering Linear \\ $(1 \rightarrow 0$)}& \parbox{17em}{\centering $\dfrac{\hbar}{2}\Omega_{24}\ket{2}\bra{4}+h.c.$}  \\ 
			\hline \parbox{4em}{\centering Circular \\ $(1 \rightarrow 0$)} & \parbox{17em}{\centering $\dfrac{\hbar}{2}\Omega_{14}\ket{1}\bra{4}+h.c.$} \\ 
			
			\hline \parbox{4em}{\centering Linear \\ $(1 \rightarrow 2$)} &  \parbox{17em}{\centering$\dfrac{\hbar}{2}(\Omega_{15}\ket{1}\bra{5}+\Omega_{26}\ket{2}\bra{6}+\Omega_{37}\ket{3}\bra{7}) $\\$+h.c.$}  \\ 
				\hline \parbox{4em}{\centering Circular \\ $(1 \rightarrow 2$)}& \parbox{17em} {\centering$\dfrac{\hbar}{2}(\Omega_{16}\ket{1}\bra{6} +\Omega_{27}\ket{2}\bra{7} +\Omega_{38}\ket{3}\bra{8}) $\\$+h.c.$}\\ 
				\hline 
		\end{tabular} 
		\end{center}

The $H_{B}$ for these transitions is described below:

\begin{center}
	\begin{tabular}{|c|c|}
		\hline  \parbox{4em}{\vspace{ .1cm}\vspace{ .1cm}}& $\bf{H_{B}}$ \\ 
		\hline \parbox{4em}{\centering Linear \\ $(1 \rightarrow 0$)} & \parbox{17em}{\vspace{ .1cm}\centering $\dfrac{\hbar}{\sqrt{2}} (\Omega^g_{L}+i\Omega^g_{t}) (\ket{1}\bra{2}+\ket{2}\bra{3})+h.c.$\vspace{ .1cm}}\\ 
		\hline \parbox{4em}{\centering Circular \\ $(1 \rightarrow 0$)} & \parbox{17em}{\vspace{ .1cm} \centering $\hbar\Omega^g_L(\ket{3}\bra{3}-\ket{1}\bra{1})+ \dfrac{\hbar}{\sqrt{2}}\Omega^g_{t}(\ket{1}\bra{2} +\ket{2}\bra{3})+h.c.$\vspace{ .1cm}}\\ 
		
		\hline \parbox{4em}{\centering Linear \\ $(1 \rightarrow 2$)} & \parbox{17em}{\vspace{ .1cm}\centering$\dfrac{\hbar}{\sqrt{2}} \{(\Omega^g_{L}+i\Omega^g_{t})(\ket{1}\bra{2}+\ket{2}\bra{3})\} +\dfrac{\hbar}{2}(\Omega^e_{L}-i\Omega^e_t) \times  (2\ket{4}\bra{5} +\sqrt{6}\ket{5}\bra{6}+\sqrt{6}\ket{6}\bra{7}   +2\ket{7}\bra{8})+h.c.$ \vspace{ .1cm}} \\ 
			\hline \parbox{4em}{\centering Circular \\ $(1 \rightarrow 2$)}& \ \parbox{17em}{\vspace{ .1cm}\centering$\hbar\Omega^g_L(\ket{3}\bra{3}-\ket{1}\bra{1})+\hbar\Omega^e_L(2\ket{8}\bra{8} $\\ $+\ket{7}\bra{7}  -\ket{5}\bra{5}-2\ket{4}\bra{4}) $\\ $ + \dfrac{\hbar}{\sqrt{2}}\Omega^g_{t}(\ket{1}\bra{2} +\ket{2}\bra{3}) +\dfrac{\hbar}{2}\Omega^e_{t}(2\ket{4}\bra{5} +\sqrt{6}\ket{5}\bra{6} $\\ $  +\sqrt{6}\ket{6}\bra{7}+2\ket{7}\bra{8})+h.c.$\vspace{ .1cm}}\\ 
			\hline 
		\end{tabular} 
		\end{center}

In the above Hamiltonians, $\Omega_{ij}$ is the Rabi frequency of laser field between levels $\ket{i}$ and $\ket{j}$, $\Omega^{g(e)}_{L}$ is Larmor frequency for ground (excited) states along longitudinal direction, and $\Omega^{g(e)}_{t}$ is transverse magnetic field for ground (excited) levels.
The Hanle resonances using the above Hamiltonians are same as that obtained by using the ADM package. This simplified model gives an insight into how the package works. Therefore, in the theoretical analysis section we have only given the results by considering all the magnetic sub-levels using the ADM package.
%

\end{document}